\documentclass[11pt]{article}
\usepackage{moriond,epsfig}
\usepackage{graphicx}
\usepackage{caption}

\def\Journal#1#2#3#4{{#1} {\bf #2}, #3 (#4)}

\def\NIMA{{\em Nucl. Instrum. Methods} A}

\def\PLB{{\em Phys. Lett.}  B}

\def\PRD{{\em Phys. Rev.} D}

\def\JHEP{{\em J. High. Phys.}}

\def\be{\begin{equation}}
\def\ee{\end{equation}}
\def\bea{\begin{eqnarray}}
\def\eea{\end{eqnarray}}

\begin{document}
\vspace*{4cm}
\title{MEASUREMENTS OF TOP QUARK PROPERTIES AT THE LHC}

\author{E. Yazgan for the ATLAS and CMS Collaborations}

\address{Department of Physics and Astrophysics, University of Ghent, \\
Proeftuinstraat 86, 9000 Gent, Belgium}

\maketitle\abstracts{
Recent measurements of top quark properties at the LHC from ATLAS and CMS experiments are presented. 
The presented results include top quark mass, dependence of top mass measurements on event kinematics, top anti-top mass difference, bottom quark content in top quark decays, W boson polarization and anomalous couplings, search for CP violation in single top events, vector boson production associated with top-antitop pairs and top polarization. 
}

\section{Introduction}
The top quark is the most massive particle known to date and has a shorter lifetime 
than the hadronizarion timescale, $1/\lambda_{QCD}$ which makes its bare quark properties directly accessible. 
The analyses presented in this note use the data collected with the ATLAS \cite{ref:atlas} 
and CMS \cite{ref:cms} 
detectors during 2011 and 2012 data taking periods with proton-proton collisions at center-of-mass energies of 7 and 8 TeV, respectively. 

\section{Top Quark Mass}\label{}
\subsection{Combined LHC Results}\label{}
Top quark mass measurements performed by the ATLAS and CMS experiments using 7 TeV data with integrated luminosities up to 4.9 fb$^{-1}$ are combined using the measurements in lepton+jets, di-lepton and all-jets final states\cite{ref:masscomb}. The combination is performed using the BLUE method \cite{ref:blue} and yields
m$_t=173.3\pm0.5$ (stat) $\pm1.3$ (syst) GeV. For the combination, the most recent  measurements have not been included yet.

\begin{figure}
\center
\begin{tabular}{cc}
\begin{minipage}[b]{.4\textwidth}
       \centering
	\psfig{figure=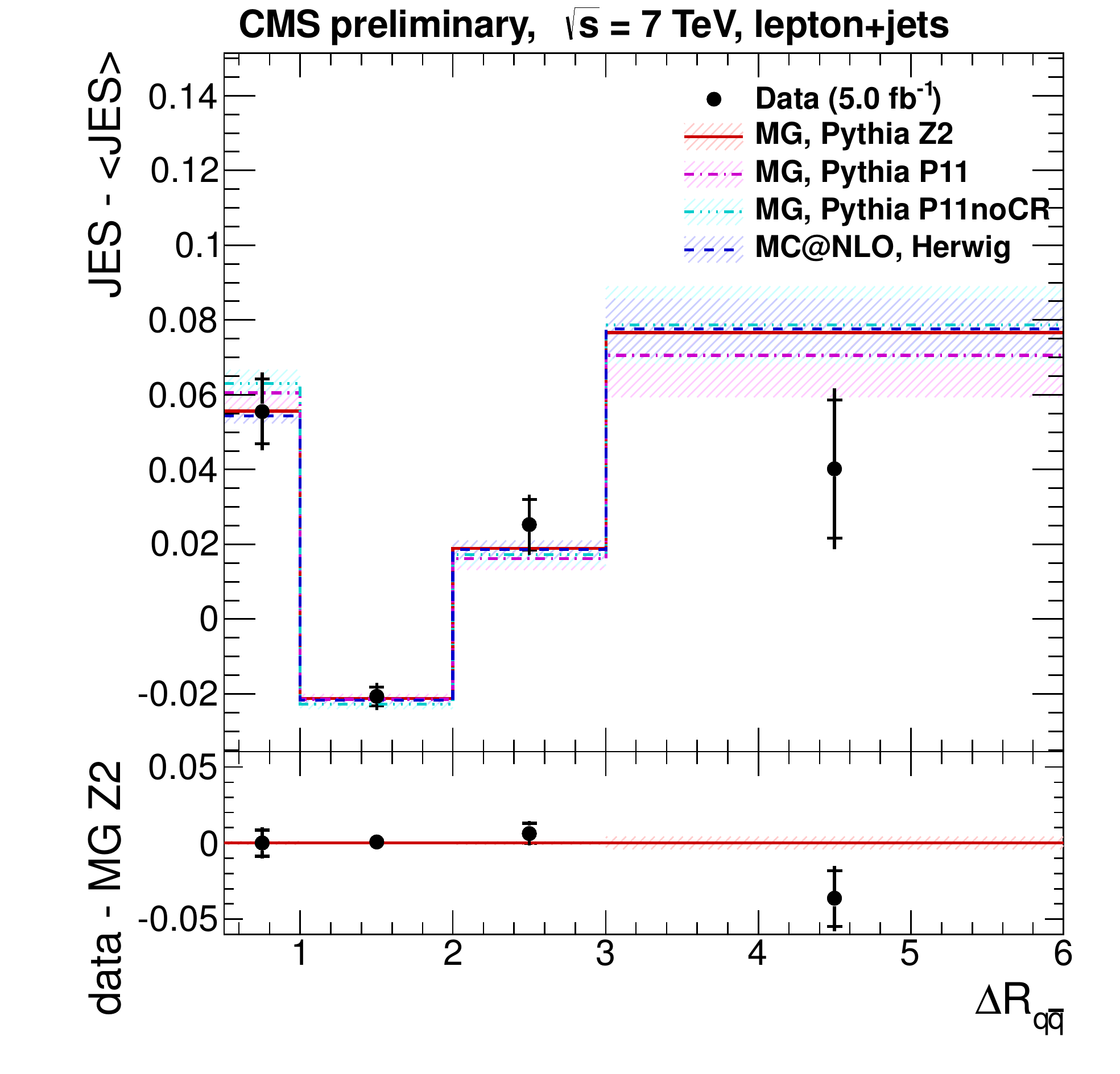, width=.9\linewidth}
\end{minipage}
&
\begin{minipage}[b]{.4\textwidth}
       \centering
\psfig{figure=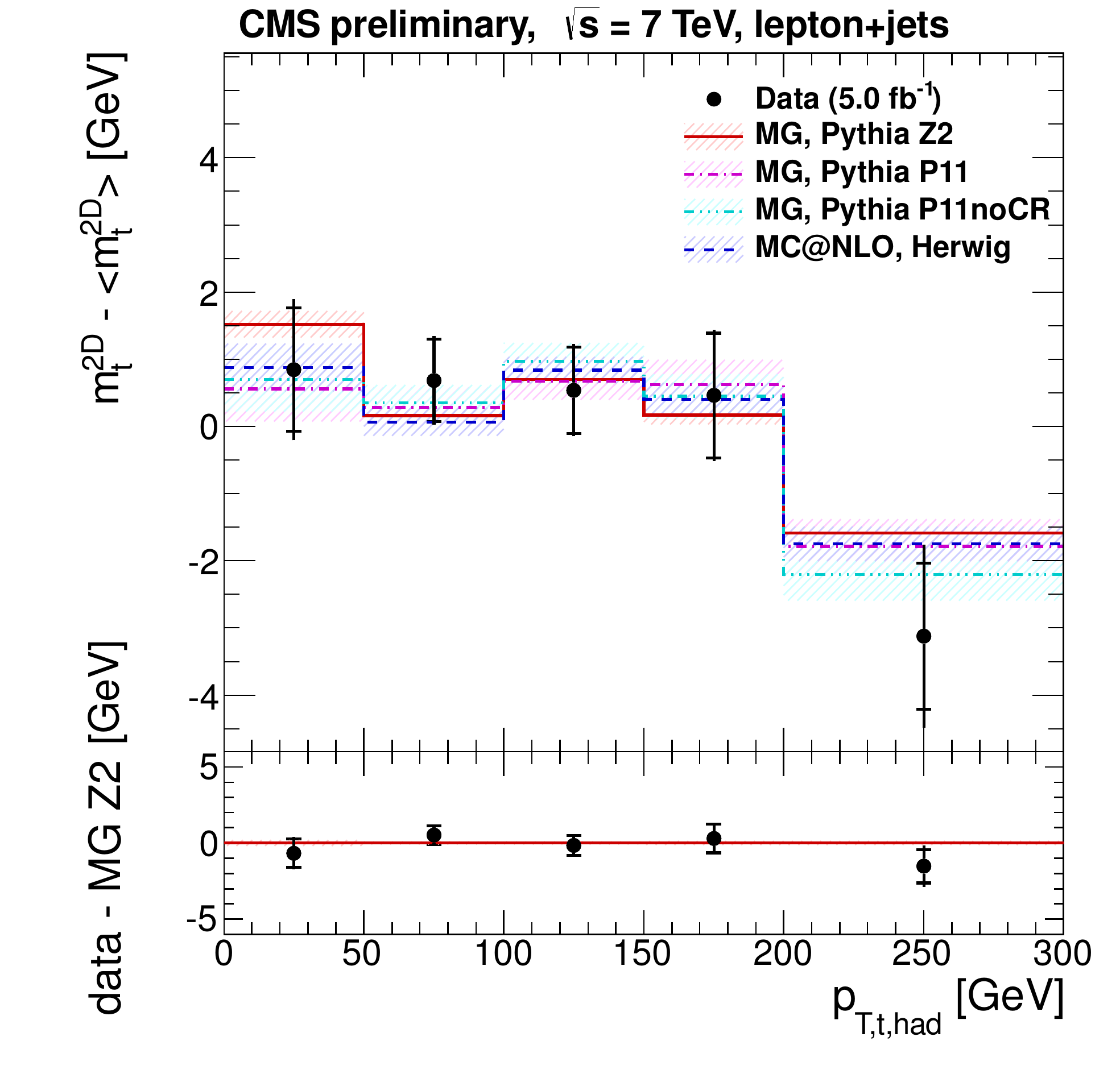,width=0.9\linewidth}
\end{minipage}
\end{tabular}
\caption{Jet energy scale vs the separation of the light-quark jets (left) and top-quark mass vs the transverse momentum of the hadronically decaying top quark (right). The systematic and statistical uncertainties are added in quadrature. The hatched areas represent the statistical uncertainties on the simulated quantities. }
\label{fig:kin}
\end{figure}

\subsection{Dependence of Top Quark Mass Measurement on Event Kinematics}\label{}
Interpretation of the top mass measurements is not straightforward when the precision becomes smaller than the top width, $\Gamma_t\sim$1 GeV. 
It is difficult to define a pole mass for an unstable and colored particle \cite{ref:colorflow1,ref:colorflow2}. 
Based on the most precise single top quark mass measurement \cite{ref:ljets}, possible kinematic biases in the measurement of top-quark mass are studied\cite{ref:topkin}. Different observables are studied which are expected to be sensitive to color reconnection effects, initial/final state radiation and b-quark kinematics (two such distributions are displayed in Figure \ref{fig:kin}). 
This study represents the first top quark mass measurement binned in kinematic observables. We conclude that with the current precision, there is no mis-modelling effect due to color reconnection, initial/final state radiation, or b-quark kinematics.

\subsection{Top Antitop Mass Difference}\label{}
The direct measurement of $\Delta m_{t}=m_{t}-m_{\overline{t}}$ tests CPT invariance. 
The measurement at $\sqrt{s}=8$ TeV \cite{ref:mttbar8} has been made using the same method utilized for the most precise single LHC top-quark mass measurement in the lepton+jets final state \cite{ref:ljets}.
The sample is divided into two samples with opposite lepton charge. The $m_{t}-m_{\overline{t}}$ difference is found to be $-272\pm196$ (stat) $\pm$ 122 (syst) MeV
consistent with 0. This represents the most precise measurement of $\Delta m_{t}$.

\section{Top Quark Couplings}\label{}
\subsection{Measurement of the Ratio ${\cal R}={\cal B}(t\rightarrow Wb)/{\cal B}(t\rightarrow Wq$)}\label{}
\begin{figure}
\center
\begin{tabular}{cc}
\begin{minipage}[b]{.4\textwidth}
       \centering
	\psfig{figure=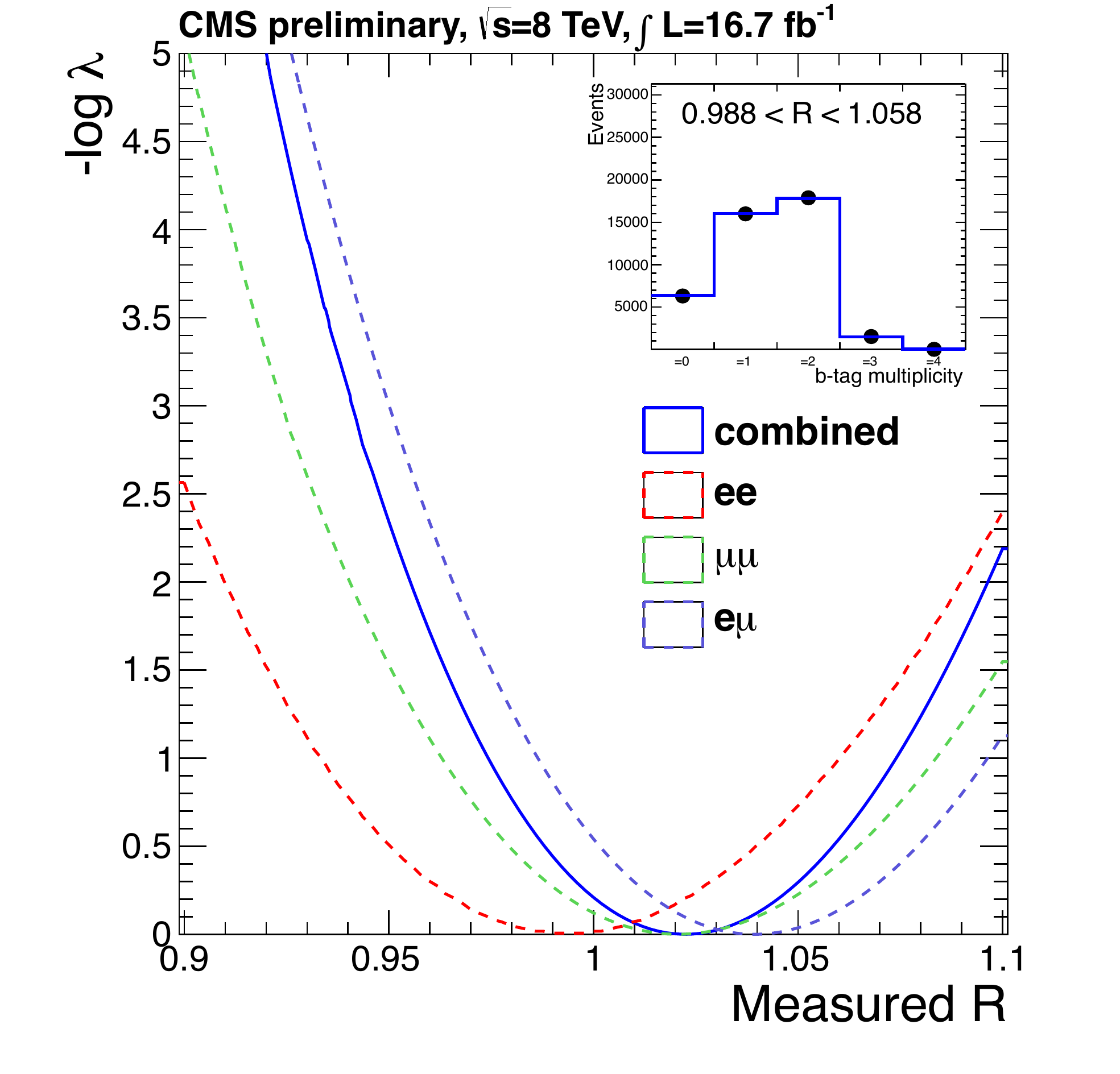, width=.9\linewidth}
	\caption{Variation of the profile likelihood used to extract R from the data. The inset shows the inclusive b-tag multiplicity distribution and the result of the fit.}
	\label{fig:rmeas}
\end{minipage}
&
\begin{minipage}[b]{.4\textwidth}
      \centering
\psfig{figure=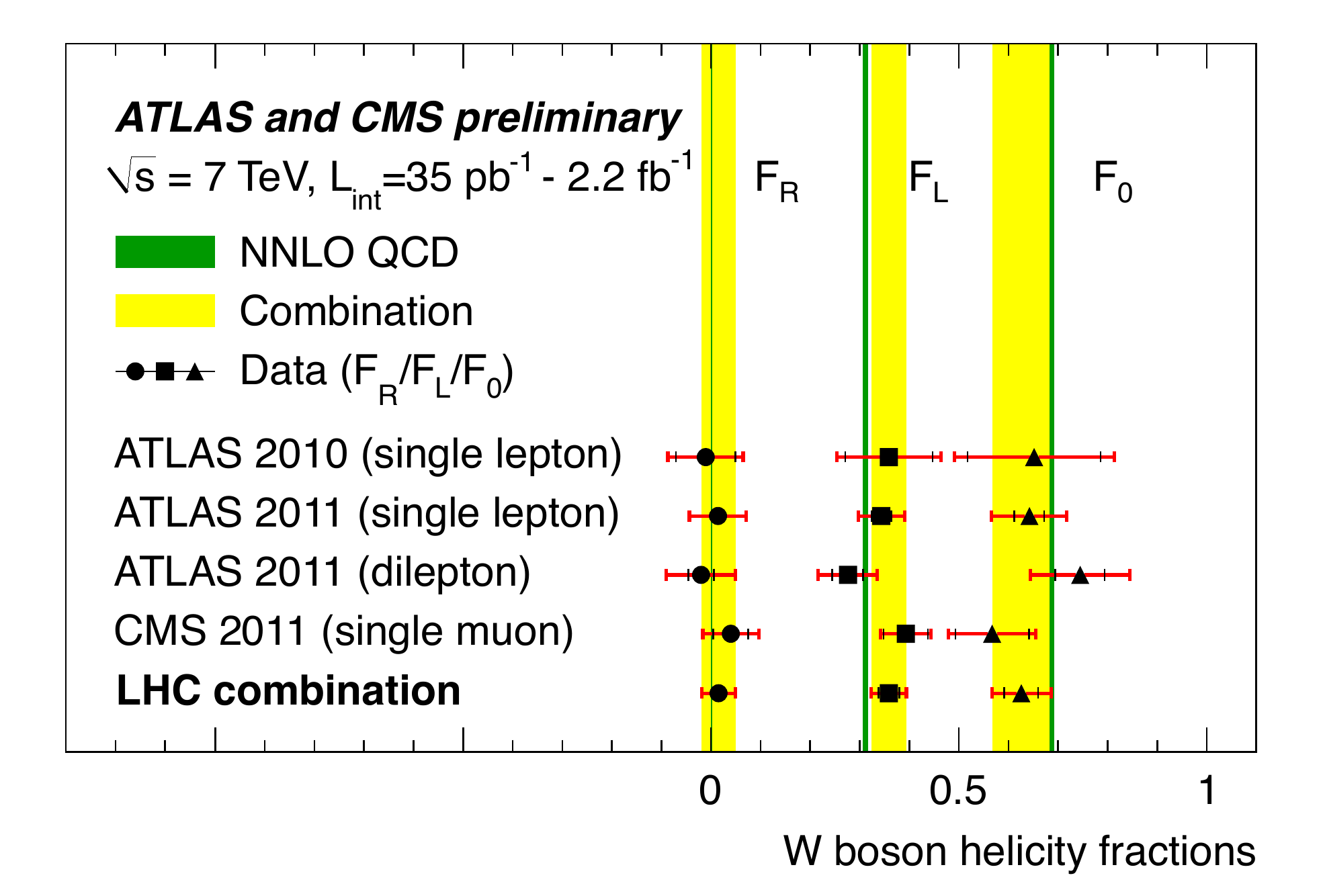,width=1.05\linewidth}
\caption{Input W helicity fraction measurements and the combination. The inner and outer error bars correspond to the statistical and the total uncertainty, respectively.}
\label{fig:whelicitylhc}
\end{minipage}
\end{tabular}
\end{figure}

In the standard model (SM), the top quark decays to a W boson and a b quark practically with a branching ratio of $\sim100$\%.
Decays to other down-type quarks are suppressed in the Cabibbo-Kobayashi-Maskawa (CKM) matrix with $|V_{tb}|>0.999>>|V_{ts}|,|V_{td}|$ under the assumptions of unitarity and exitence of three quark generations.  A measurement of $|V_{tb}|$ can be made by measuring the branching fraction ratio ${\cal R}={\cal B}(t\rightarrow Wb)/{\cal B}(t\rightarrow Wq$). 
The measurement of ${\cal R}$ is perfomed in the $t\overline{t}$ di-lepton final state measured using 8 TeV data \cite{ref:rmeas}. 
${\cal R}$ is extracted from a profile likelihood fit to data-drived analytic probability models of signal purity, number of reconstructed tops in different jet categories and number of b-tags as 1.023$^{+0.036}_{-0.034}$ (stat+syst) without any constraints in the fit (see Figure \ref{fig:rmeas}).
If the physical condition of ${\cal R}\leq1$ is imposed, then a lower limit for ${\cal R}$ is derived to be ${\cal R}>0.945$ at 95\% C.L. 
With the assumption of CKM unitarity and existence of three generations, the measurement can be converted to a measurement of $|V_{tb}|$. The CKM matrix element $|V_{tb}|$ is measured as $1.011^{+0.018}_{-0.017}$ and the lower limit is derived to be $|V_{tb}|>0.972$ at 95\% C.L.  The results obtained for ${\cal R}$ and $|V_{tb}|$ are consistent with the SM predictions and are the most precise measurement of ${\cal R}$ and the most stringent direct lower limit on $|V_{tb}|$. 

\subsection{W Boson Polarization and Anomalous Couplings}\label{}
The top quark partial with can be parametrized in terms of left-handed (F$_L$), right-handed ($F_R$), longitudinal ($F_0$) W boson helity fractions and 
the $\theta^*$ angle between the momentum of the d-type fermion in the W boson rest frame and the momentum of the W boson in the top quark rest frame. 
Measurements of helicity fractions probe the anomalous couplings, $g_L$ and $g_R$. 
ATLAS and CMS  made measurements of W boson helicities in $t\bar{t}$ events in dilepton and lepton+jets final states \cite{atlaswhel1,cmswhel1,cmswhel2} and combined the W helicity measurements \cite{ref:whelcomb} at $\sqrt{s}=$ 7 TeV. 
All measurements are found to be consistent with each other and with SM predictions at NNLO QCD (See Figure \ref{fig:whelicitylhc}). 

Measurement of W Helicity fractions is also performed in single-top topologies \cite{ref:whelsingletop} studying $\mu+jets$ final state at $\sqrt{s}$ = 7 and 8 TeV. The helicity fractions are obtained from likelihoods with reweighted signals including all processes involving the top quark. 
The measurements performed at $\sqrt{s}$ = 7 and 8 TeV are combined yielding $F_0=0.713\pm0.114$ (stat)$\pm0.023$ (syst), $F_L=0.293\pm0.069$ (stat)$\pm0.030$ (syst) and $F_R=-0.006\pm0.057$ (stat)$\pm0.027$ (syst).  The results are found to be consistent with the SM and measurements in the $t\bar{t}$ channels.
All the W helicity measurements discussed above are also used to set exclusion limits on the real part of the anomalous couplings. 

\subsection{Search for CP Violation in Single Top Events}\label{}
The angle $\theta^*$ is not sensitive to the complex phases of anomalous coplings. Non-zero complex phases might indicate a CP-violating component in the top quark decay. 
In the $t$-channel single top production, the top quarks are produced with a high degree of polarization along the spectator quark direction. 
The vector, $\vec{N}$, normal to the top polarization and W (or b) directions in the helicity frame can be used to define forward-backward asymmetry ($A_{FB}^N$) which is proportional to the imaginary part of $g_R$.
Using the unfolded distributions of $\cos\theta^N$, ATLAS collaboration measured the forward-backward asymmetry at the parton level \cite{ref:cpviol} in the $lepton+jets$ channels at $\sqrt{s}=7$ TeV as $A_{FB}^N=0.031\pm0.065$ (stat)$^{+0.029}_{-0.031}$ (syst). 
Assuming a polarization of $P=0.9$\cite{ref:st_pol1,ref:st_pol2}, the measurement is converted into limits on $\Im(g_R)=[-0.20,0.30]$ at the 95\% C.L. This represents the first limit on $\Im(g_R)$ and the results are consistent with the SM predictions. 

\subsection{Vector Boson Production Associated with Top-Antitop Pairs}\label{}
Studying vector boson production in association with $t\bar{t}$ pairs provides a test of the SM top quark-vector boson coupling. Measurement of these couplings are important for new physics searches and also the discovery of the Higgs boson in $t\bar{t}H$ process. The first cross section measurements of the $t\bar{t}V$ processes are done at the LHC \cite{ref:ttvatlas,ref:ttvcms} using the same-sign dilepton for $t\bar{t}W$ and $t\bar{t}Z$ and trilepton signature for $t\bar{t}Z$ process at $\sqrt{s}=7$ TeV. The cross section for $t\bar{t}V$ is measured by CMS to be $0.43^{+0.17}_{-0.15} (stat)^{+0.09}_{-0.07} (syst)$ pb and $t\bar{t}Z$ to be $0.28^{+0.14}_{-0.11} (stat)^{+0.06}_{-0.03} (syst)$ pb. Both ATLAS and CMS measurements are consistent with the SM NLO calculations \cite{ref:cambell12,ref:garzelli12}.
The $t\bar{t}Z$ measurements represent the first direct measurements of top quark-Z boson coupling. 

\section{Top Polarization}\label{}
Top quarks are produced unpolarized in the $t\bar{t}$ process. This is verified by ATLAS in the lepton+jets \cite{ref:ATLAS-CONF-2012-133} and by CMS in the dilepton final state \cite{ref:CMS-PAS-TOP-12-016}. The distribution of the polar angle is proportional to $1+\alpha_ip\cos\theta_i$ in the parent top rest frame.
ATLAS used templates with different $\alpha\times p$ values to fit to the reconstructed $\cos\theta_l$ distribution and extracted the fraction of positively polarized top quarks, $f=0.470\pm0.009$ (stat)$^{+0.023}_{-0.032}$ (syst) compatible with the SM expectation.

\section{Conclusions}
Measurements of top quark properties by ATLAS and CMS at the LHC are providing thorough tests of the SM. 
All top quark properties measurements at the LHC show good agreement with the SM predictions.

\end{document}